# A Machine Learning Framework for the Prediction of Grain Boundary Segregation in Chemically Complex Environments

Doruk Aksoy [1], Jian Luo [2], Penghui Cao [1,3], Timothy J. Rupert [1,3,4,5,*]

[1] Department of Materials Science and Engineering, University of California, Irvine, CA 92697, USA

[2] Department of NanoEngineering, University of California San Diego, La Jolla, CA 92093, USA

[3] Department of Mechanical and Aerospace Engineering, University of California, Irvine, CA 92697, USA

[4] Hopkins Extreme Materials Institute, Johns Hopkins University, Baltimore, MD 21218, USA

[5] Department of Materials Science and Engineering, Johns Hopkins University, Baltimore, MD 21218, USA

[*] Corresponding author: tim.rupert@jhu.edu

**ABSTRACT**

The discovery of complex concentrated alloys has unveiled materials with diverse atomic environments, prompting the exploration of solute segregation beyond dilute alloys. However, the vast number of possible elemental interactions means a computationally prohibitive number of simulations are needed for comprehensive segregation energy spectrum analysis. Data-driven methods offer promising solutions for overcoming such limitations for modeling segregation in such chemically complex environments, and are employed in this study to understand segregation behavior of a refractory complex concentrated alloy, NbMoTaW. A flexible methodology is developed that uses composable computational modules, with different arrangements of these modules employed to obtain site availabilities at absolute zero and the corresponding density of states beyond the dilute limit, resulting in an extremely large dataset containing 10 million data points. The artificial neural network developed here can rely solely on descriptions of local atomic environments to predict behavior at the dilute limit with very small errors, while the addition of negative segregation instance classification allows any solute concentration from zero up to the equiatomic concentration for ternary or quaternary alloys to be modeled at room temperature. The machine learning model thus achieves a significant speed advantage over traditional atomistic simulations, being four orders of magnitude faster, while only experiencing a minimal reduction in accuracy. This efficiency presents a powerful tool for rapid microstructural and interfacial design in unseen domains. Scientifically, our approach reveals a transition in the segregation behavior of Mo from unfavorable in simple systems to favorable in complex environments. Additionally, increasing solute concentration was observed to cause anti-segregation sites to begin to fill, challenging conventional understanding and highlighting the complexity of segregation dynamics in chemically complex environments.

# 1. INTRODUCTION

The microstructural design of materials is crucial for applications that demand high performance and/or operation under extreme conditions [1]. The design challenge lies in understanding and manipulating behaviors at the atomic level, where the arrangement of atoms and their interactions dictate the resultant macroscopic properties of the material. In this context, chemically complex environments (CCEs), which include solute-solute interactions and thermal effects, present a unique opportunity for investigation, with complex concentrated alloys (CCAs) or high entropy alloys (HEAs) serving as examples. Among materials exhibiting CCEs, refractory complex concentrated alloys (RCCAs) stand out as they are not only superior in high-temperature applications compared to Ni-based superalloys [2] but also offer a rich landscape for studying phenomena such as chemical short-range ordering (CSRO) and solute clustering behavior [3–5]. The equimolar NbMoTaW RCCA, in particular, has been extensively studied due to its exceptional high-temperature strength and its ability to maintain a stable single-phase at elevated temperatures [6,7].

Prior study of NbMoTaW and other RCCAs has shed light on the interplay between CSRO and interface structure on grain boundary segregation [8–11], an important behavior that dictates mechanical properties such as strength, ductility, and fracture. The findings of these studies highlight the complexity of microstructural modification in RCCAs, characterized by both appearance of distinct CSRO patterning [12,13] and the competition among multiple solutes for the same lattice sites [14–16]. This competition leads to a wide array of interesting segregation phenomena, with chemical patterning amplified by variations in site occupancy probability and the consequent impact on embrittlement at grain boundary sites [17,18]. These variations are attributed to differences in interface structure and interactions with neighboring boundaries within the polycrystalline grain boundary network [14,15]. The non-uniform distribution of segregation energies with such a network results in a complex pattern of site availability for segregation. Furthermore, prior investigations have demonstrated that in the transition regions from interfaces to bulk sites, the compositional variation is notably affected by the presence of interfaces [19,20], with distinct structural ordering preferences and extended segregation zones being observed. This last finding indicates that segregation behavior is not confined to the immediate interface but instead extends into adjacent zones, resulting in areas that do not exhibit characteristics typical of either pure interface or bulk behavior. This diversity of interfacial behaviors is magnified in CCEs due to the increased number of principal elements, making the prediction and control of segregation behaviors at interfaces a considerable challenge.

Early work in RCCAs was mostly focused on equiatomic systems [7,21], which represents only a single composition out of a huge possible set, leaving many off-equiatomic systems largely explored. The increased complexity associated with CCEs can be understood by thinking about chemical interactions between intra-species (between the same species) and inter-species (between different species) pairs. In a CCE, the interactions between a binary pair (e.g., Nb-Mo) when only these two elements are present in a system are not chemically equivalent to the same pair when additional elements are present. Therefore, to understand the effect of solute segregation, these interactions must be treated as a large collection of A-B type interactions, where B is a single element as solute (e.g., Ta), and A is the "base" material. For example, A could be a single element (e.g., Mo), or a mixture of elements (e.g., $Nb_{0.5}Mo_{0.5}$ or $Nb_{0.34}Mo_{0.33}Ta_{0.33}$). This distinction enables the study of different concentrations of B as possible solute in various A bases. Furthermore, since the various chemical environments any given dopant can be surrounded by are also accounted for in a multi-element alloy base, the segregation behavior in the non-dilute case is also captured by such a framework. In a quaternary system, there are 32 possible A-B combinations (4 intra-species and 28 inter-species). Inter-species interactions can be divided into three subparts, with A comprising of a single element (A-B representing a binary system), or A as a multi-element alloy (A-B representing a ternary or quaternary system). For an A-B pair below the dilute limit, the number of thermodynamically different interfacial sites can be huge depending on the size of the model and ratio of interface to bulk sites. For instance, the Wagih and Schuh study [22] had to treat ~415,000 interfacial sites for the Al-Mg binary system with a polycrystal of 36 nm edge length and an atomic site fraction of ~15%. The edge length refers to the dimension of the cubic simulation cell used to construct the polycrystal, which contains 96 randomly oriented grains. The size of the polycrystal was selected through comparisons of disorientation distributions and effective segregation energies to ensure a representative sample of the material's grain boundary network. If a similar direct segregation site simulation was employed for a quaternary alloy, restricted to the dilute limit, the number of simulations to obtain the spectrum of segregation energies at absolute zero needs to be multiplied by the number of inter-species interactions, resulting in approximately 11 million simulations. For CCEs with a broad range of concentrations, this number is virtually infinite, but for a more practical application the following scenario can be considered. Assuming an equimolar base at room temperature, if the solute concentration is varied in 0.25 at.% increments up to equimolar composition, the number of concentrations to be considered are 200 for binaries, 135 for ternaries and 100 for quaternary, resulting in ~4.8 billion simulations. Considering that this calculation is for an equimolar base at a certain temperature, changing any of these parameters will increase the number of simulations exponentially.

Data-driven methods offer a promising avenue for constructing structure-property models that can overcome the limitations of current predictive tools arising from the combinatorial increase problem. These methods, which have been leveraged in past research efforts to perform tasks such as phase classification [23–29], search for ideal structural descriptors [30–39], and study of metastable structures and synthetic microstructure generation [40–42], can potentially unravel the intricate patterns of segregation behavior and microstructural properties across a broad compositional landscape. While previous studies have employed neural networks and other advanced algorithms to model segregation patterns [43–48], accurately predicting interfacial segregation in the CCEs at non-zero temperatures remains a challenge.

The present work introduces a methodology capable of predicting interfacial segregation within CCEs, employing only the descriptions of local atomic environments. The thermodynamic framework for binary alloy systems faces challenges in CCEs due to the complex interactions among multiple elements. This study first introduces an expansion to such a framework for CCEs, then presents a flexible methodology using composable computational modules. These modules include the generation of data through atomistic simulations and the application of optimized machine learning (ML) models, including artificial neural networks (ANNs) and gradient boosted decision trees. These modules can be rearranged depending on the desired output data, to study segregation in CCEs and accommodate various compositions and temperatures. Tailored computational pipelines are proposed for a few example applications, accelerating the transition towards ML-driven predictions in CCEs. Utilizing these pipelines, diverse trends in segregation energies are observed, highlighting the complexity of predicting interfacial segregation in CCAs. For instance, a transition in the segregation behavior of Mo is observed, shifting from unfavorable segregation in simple systems to increasingly favorable segregation as the alloy compositional complexity grows. Moreover, our methodology uncovers the complex dynamics of solute-solute interactions and thermal effects in CCEs, revealing that with increasing solute concentration anti-segregation sites begin to fill up, leading to a more complex segregation landscape than previously understood. This challenges the expectation of a narrowing trend in segregation behavior seen in binary alloys, suggesting a revised mechanism where even energetically favorable sites are not always occupied and highlighting the intricate interplay of solute concentrations as well as their impact on material properties in CCEs. This study not only contributes to the field by offering a robust tool for researchers to explore the vast compositional space of CCEs but also aids in the design and development of CCEs with tailored microstructural attributes of new materials for the technological demands of the future.

## 2. METHODS

**Thermodynamic Framework**

A typical thermodynamic framework for interfacial solute segregation in binary alloys at the dilute limit assumes non-interacting solutes, which can be captured by the Gibbs free energy difference between bulk and interfacial sites. This relationship can be described by the McLean isotherm [49]:

$$\Delta E_{seg}^{int} = E_b^{int} - E_{bulk} \qquad (1)$$

where $\Delta E_{seg}^{int}$ represents the interfacial segregation energy, $E_b^{int}$ is the binding energy at a solute-occupied interfacial site, and $E_{bulk}$ is the binding energy within the bulk matrix. This model simplifies the segregation energy to a single value, or a single site type, independent of solute concentration and temperature. However, experimental evidence has shown that this single-value approach does not accurately describe segregation behavior [50]. Consequently, the White-Stein-Coghlan model was introduced to extend this model to a multiple site-type model, which accounts for the spectral distribution of segregation energies at various interface sites, which describes site availabilities at absolute zero temperature [50]. In this context, a "type" refers to the distinct atomic configurations at the grain boundary, each with unique binding energies that influence segregation behavior. For each type of interfacial atom $i$, the model defines a site-specific concentration, $X_i^{int}$, as a function of the bulk solute concentration, $X^{bulk}$, and the site-specific segregation energy, $\Delta E_{i,seg}^{int}$:

$$X_i^{int} = \frac{X^{bulk} \exp\left(-\frac{\Delta E_{i,seg}^{int}}{kT}\right)}{1 - X^{bulk} + X^{bulk} \exp\left(-\frac{\Delta E_{i,seg}^{int}}{kT}\right)} \tag{2}$$

where $k$ is the Boltzmann constant and $T$ is the absolute temperature. This model employs Fermi-Dirac statistics, which dictates that each interfacial site can be occupied by at most one solute atom, with the occupancy being temperature-dependent. The probability of site occupancy is then given by:

$$P_i^{int} = F_i^{int} X_i^{int} \tag{3}$$

where $F_i^{int}$ represents the density of states (DOS) for the interfacial sites of type $i$, calculated by the ratio of the number of interface sites with segregation energy $\Delta E_{i,seg}^{int}$ to the total number of interfacial sites:

$$F_i^{int} = \frac{N_i^{int}}{\sum_i N_i^{int}} \tag{4}$$

While the thermodynamic framework described above offers a robust foundation for understanding solute segregation in binary alloy systems, its application to CCEs has significant challenges. The traditional binary approach does not capture the synergistic and antagonistic effects between multiple species in CCEs. Therefore, to accurately represent the inherently dynamic and localized nature of interfacial segregation within these multi-element systems, the model must integrate the combined influence of various solute atoms on segregation behavior, deviating from the ideal McLean model. Furthermore, in binary systems, reference states are well-defined as the pure base element, facilitating straightforward calculations of segregation energies and site-specific behaviors. However, in alloys composed of more than two elements, such as CCAs, the definition of reference states becomes non-trivial due to the multitude of elemental interactions. Since a clear reference state cannot be defined unlike the binary case, the White-Stein-Coghlan model requires adaptations to reflect the chemical complexity of CCEs. To accomplish this, we adopt a flexible methodology that involves creating pipelines tailored for different applications through the use of composable computational modules shown in Fig. 1. These modules correspond to different procedures and tools, such as sample preparation, atomistic simulations (A1-A3), preprocessing, and ML techniques (M1 and M2). Different arrangements of these modules, referred to as pipelines, can be used to obtain site availabilities at absolute zero as well as DOSs in CCEs with or without atomistic simulations. In the next sections, these modules will be introduced and then the pipelines and their corresponding module arrangements for different applications will be described.

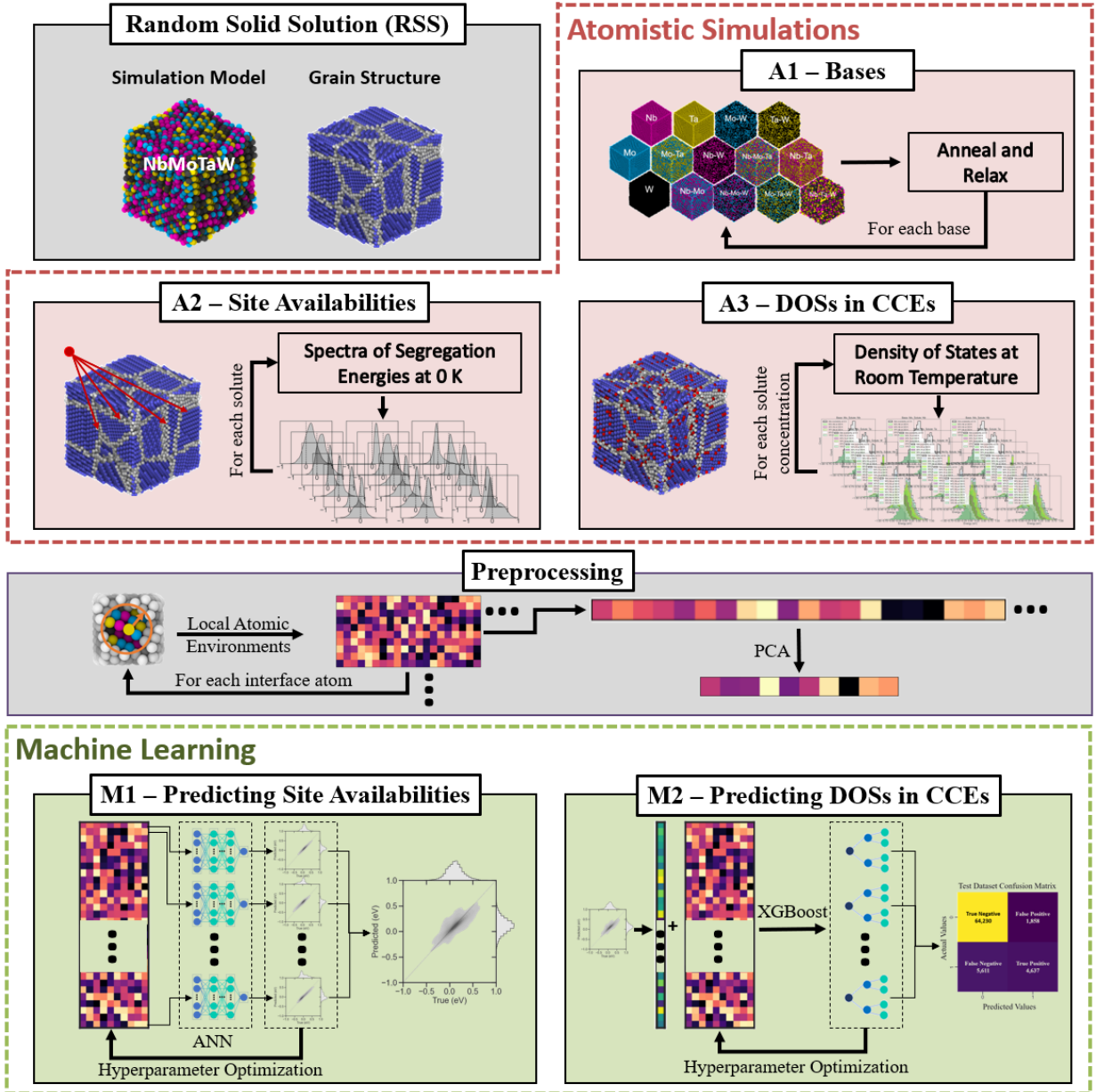

Fig. 1. Illustration of composable computational modules. After the selection of a material system, random solid solution models (RSS module) are annealed and relaxed (A1 module), while site availabilities (A2 module) and density of states at room temperature (A3 module) are calculated via atomistic simulations. The preprocessing module involves principal component analysis to reduce the dimensionality of vectorized local atomic environments. The methodology utilizes artificial neural networks (M1 module) and gradient-boosted decision trees through XGBoost (M2 module) for predictions, addressing the complexity of multi-elemental

**interactions and various conditions. The dashed lines enclosing the modules represent the type of computational method used.**

**Sample Preparation and Bases**

The first module in Fig. 1, the regular solid solution (RSS) module, involves the selection of a material system (NbMoTaW for this work) and the creation of the simulation model. Computational models of polycrystalline NbMoTaW were prepared using Atomsk [51], an open-source program for creating atomic-scale models. A $50 \times 50 \times 50$ Å$^3$ model comprising four randomly oriented grains was generated. This model and grain size selection was based on the assumption, substantiated by findings in binary systems [52], that in a quaternary alloy, the chosen dimension is sufficiently large to ensure that segregation behavior remains relatively unaffected by grain size, owing to the reduced fraction of grain boundaries. The choice of model size was validated through a comparative analysis with a larger $150 \times 150 \times 150$ Å$^3$ model for the equiatomic quaternary system, demonstrating the representativeness of the smaller model in capturing segregation energy distributions. Detailed discussion and supporting data are provided in Supplementary Note 1.

In the second module, A1, the simulation model created in the RSS module is utilized to create different bases, 14 in total for this study, which represent all possible equiatomic combinations. These bases (A) can be either single element (Nb, Mo, Ta, and W) or multi-element equiatomic alloys (e.g., NbMo and MoTaW). The construction process involved the creation of a polycrystal with one element (Nb) initially, preserving the same orientation while creating other bases, resulting in 14 bases that match the corresponding average lattice parameters of the alloying system. These bases serve as the starting point for subsequent simulations, so that solute species (B) can be substituted.

In the current study, a combination of molecular statics and hybrid Monte Carlo/molecular dynamics (MC/MD) simulations were employed with the Large-scale Atomic/Molecular Massively Parallel Simulator (LAMMPS) software package [53]. OVITO was used for atomic visualizations [54], while data visualization was accomplished using Python libraries Matplotlib [55] and Seaborn [56]. The interatomic potential used in this work [57] was derived from a machine learning interatomic potential based on the Moment Tensor Potential methodology [58]. This approach, involving invariant vectors to represent local atomic environments, was validated against density functional theory simulations by Yin et al. [59], showing good agreement in properties such as melting temperature, unstable stacking fault energy, and elastic constants, as well as prediction of the expected chemical short-range ordering behavior [57].

Composition adjustments were made by randomly substituting atoms of different types until the desired composition was achieved. The system was then heated to room temperature under a canonical (NVT) ensemble for 250 ps, followed by controlled cooling to 0 K at a rate of 3 K/ps, before finally a conjugate gradient energy minimization procedure was applied. This process yielded equimolar random solid solution alloy systems comprised of one, two, and three constituent elements. This procedure was replicated across all 14 bases that have been established for this study. A key advantage of this approach lies in its flexibility; the same base can be repurposed to study different solutes, such as substituting Ta for W as the potential segregating species in NbMo, or to examine varying solute concentrations. Furthermore, this method allows for the examination of these configurations at various temperatures. This strategy enables a categorization of elemental interactions and paves the way for investigations of segregation behavior in different combinations of bases and solutes.

**Site Availabilities at Absolute Zero**

In the A2 module, site-specific segregation energies for compatible solute atoms were determined for each base configuration. The process involved substituting a solute atom at interface sites sequentially, followed by an energy minimization to relax the model. The interfacial segregation energy for each site was derived from the difference in energy after substituting the solute atom at an interface site compared to a bulk site. Given the heterogeneous distribution of principal elements in a complex concentrated alloy, calculating a representative bulk binding energy required first establishing a distribution of bulk binding energies for each elemental combination. These energy distributions are shown in Supplementary Note 2. Considering the entire spectrum associated with bulk energies would complicate the analysis, as it introduces two distributions from which to calculate segregation energies. However, a simplification was made because the energies at the fringes of the distributions, shown in Fig. S2, do not significantly affect segregation behavior, as they are less frequent, and those closer to the mode have a lower energy difference compared to the mode (around 0.3 eV), which is detailed in Supplementary Note 2. Therefore, the mode of these distributions was selected as the representative bulk atomic configuration, which corresponds to the most frequent bulk site type in the system. For example, in a $Nb_{0.5}Mo_{0.5}$ base, the representative bulk atom for Nb was identified, and the system's energy is calculated following the substitution of a W atom at this bulk site after its subsequent relaxation. A similar procedure was repeated for Mo as a bulk site. The interface site substitution spectrum is thus found by comparing this representative bulk binding energy to the local site energy for each interface site to calculate the segregation energy. This approach was replicated for each interface site, alternative solutes, and across all base configurations, resulting in the generation of site availability distributions at absolute zero. Strict energy minimization procedures with very low energy tolerance ($10^{-18}$) were employed to determine the energies related to segregation behavior, as performed by Wagih and Schuh [22].

**Chemically Complex Environments**

In the A3 module, solute atoms were first randomly substituted into the host matrix without distinguishing between interface and bulk sites, until the desired alloy composition was achieved. Hybrid Monte Carlo and molecular dynamics (MC/MD) simulations at room temperature were then conducted to calculate site occupancies. The simulations involved 1000 MC steps for every MD step, tracking the change in atomic fractions, with solutes being assigned to random sites and a Metropolis algorithm used to determine if site jumping occurs. The MD portion was performed under an isothermal–isobaric (NPT) ensemble. In contrast to the A2 module simulations which disregarded composition effects by only probing the physics of segregation at the dilute limit, these MC/MD calculations inherently incorporate solute-solute interactions. To effectively capture the atomic environments in CCEs and translate to a format conducive for machine learning analysis, a binary classification was applied to each interface site within the equilibrated configuration after MC/MD simulations, with a label of '1' used for occupation by a solute atom and a label of '0' used otherwise. Binary designation of these states forms an index for evaluating extent of negative segregation, the redistribution process where segregated solutes segregate back into the bulk, revealing deviations from initially anticipated segregation patterns influenced strictly by preferential site occupancy.

The binary descriptions of the segregation states obtained through this matching procedure are referred to as the MC/MD states. This data was acquired for all intended solute concentrations, which was investigated in 5 at.% increments at room temperature. As demonstrated previously, simulating smaller increments exponentially increases the computational overhead, which also substantiates the need for ML methods. To condense the simulation workload, base atoms were presumed to maintain an equimolar concentration, with only the solute concentrations varying. As an example, for a ternary system targeted at a 30 at.% solute concentration of Ta, the two base elements were adjusted to 35 at.% each.

**Preprocessing**

The initial step related to data preprocessing for machine learning algorithm training requires vectorization of the local atomic environments through structural descriptors that satisfy certain properties: (1) permutation invariance, (2) rotation invariance, and (3) translation invariance [60]. Local atomic environment descriptors previously reported in the literature include atom-centered symmetry functions [61], the coefficients of the bispectrum of the atomic neighbor density functions [62], rotationally covariant tensors [58] for developing Moment Tensor Potentials [57,60], and the smooth-overlap of atomic positions (SOAP). In this work, SOAP vectors were selected, since these vectors ensure that the local atomic environments were represented in a manner that is consistent and invariant to permutation, rotation, and translation, surpassing the capabilities of other conventional structural descriptors (e.g., atomic volumes) that are typically limited to first nearest neighbors [63]. SOAP vectors for these random solid solution models were calculated using Atomic Simulation Environment (ASE) and QUIP libraries [64,65], employing a method similar to that described in Ref. [66]. Moreover, the use of SOAP vectors supports effective dimensionality reduction, allowing for a computationally efficient approach while retaining critical information about the atomic environments. Principal component analysis (PCA) was used for dimensionality reduction, reducing the number of SOAP vectors, thereby decreasing the time and computational demand for the training phase. The PCA analysis is visualized in Fig. 2, where individual bars indicate the variance explained by each principal component, while the line shows the cumulative variance. The first ten components were used to represent the local atomic environments, based on the trade-off between computational efficiency and maintaining the integrity of the original high-dimensional data, as these 10 components account for 99.5% of the explained variance within the dataset.

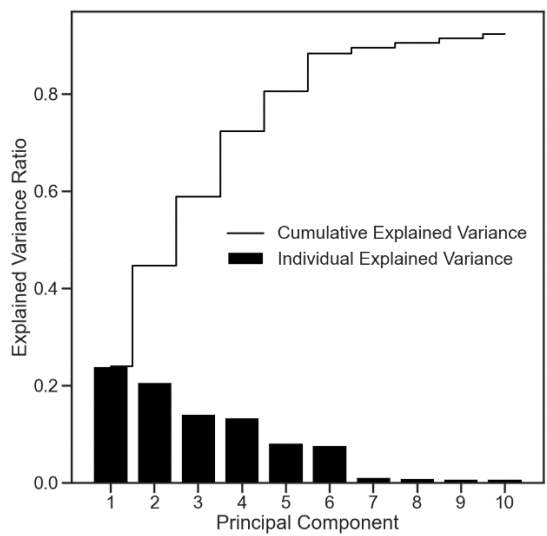

**Fig. 2. Principal component analysis of SOAP vectors for local atomic environment representation. Individual explained variance by each principal component and the cumulative explained variance are shown as filled black boxes and a black line, respectively. The first ten components capture 99.5% of total variance, effectively representing local atomic structures for subsequent machine learning analysis.**

Following the vectorization of local atomic environments, the datasets for regression or classification tasks were prepared. The resulting datasets contained over 10 million data points. The datasets were then standardized using a feature scaler to remove the mean and scale to unit variance. For the regression task in module M1, SOAP vectors served as inputs with segregation energies as targets, and the dataset was shuffled and divided into train, validation, and test datasets with percentages of 70%, 20%, and 10%, respectively. For the classification task in M2, both SOAP vectors and segregation energies were used as inputs, contingent on the solute concentrations and temperatures, while the targets were designated as the MC/MD states. The dataset, after shuffling, was split into training and test datasets with percentages of 80% and 20%, respectively, employing stratification to maintain class proportionality.

**Predictive Modeling with Machine Learning**

The M1 module aimed to link the interfacial segregation behaviors at the dilute limit to local atomic environments through the application of ANNs. Following preprocessing, the neural network architecture was designed for regression analysis. The metric to assess the regression model's performance was selected as the mean squared error (MSE), which was favored over mean absolute error (MAE) due to the spectral nature of segregation energies and the ability of MSE to prevent the cancellation of positive and negative values associated with segregation and anti-segregation sites [67]. Additionally, MSE amplifies larger errors, which are indicative of the extremities found in the skew-normal distribution of segregation energies in polycrystals [15].

The regression analysis employed Python libraries Tensorflow and Keras. The constructed neural network, a sequential model typically used for nonlinear complex problems, included batch normalization and dropout at a rate of 0.2 to enhance model performance and reduce overfitting. The hidden layers used the rectified linear unit (ReLU) activation function, while the output layer utilized a linear activation. MSE served as the loss function, with the Adam optimizer being employed. Training included early stopping and model checkpoint callbacks for enhanced efficiency. Hyperparameter optimization was conducted initially through a randomized search with 20 iterations, followed by a more targeted grid search. Parameters such as the number of hidden layers, number of neurons, learning rate, and optimizer type were tuned to improve model performance. Parameters related to SOAP vectors were also optimized during the hyperparameter tuning procedure, where a selection of 12 basis functions for angular and 12 for spherical interactions was found to yield the best accuracy with a 1 Å Gaussian smearing width of atom density within a 6 Å cutoff. PCA was then applied for dimensionality reduction to 10 principal components, facilitating faster training and prediction times with reduced computational resources at a slight trade-off in accuracy (~2-3% depending on the case), resulting in a different set of SOAP descriptors for each base material due to the many-body nature of SOAP formality. Cross-validation involving a split of the dataset into 5 parts was used during the hyperparameter tuning process to validate the robustness of the hyperparameter combinations. The final hyperparameters of the optimized model are presented in Supplementary Note 3.

In the second module of the ML block, designated as M2, a machine learning model was trained that targeted the binary classification of the MC/MD states. Extreme Gradient Boosting (XGBoost) [68], an optimized decision tree-based model that employs gradient boosting, was chosen for this task. Decision trees, which facilitate recursive partitioning, adapt to nonlinear data structures, while gradient boosting enhances model accuracy by combining weak learner predictions sequentially. Additionally, XGBoost was selected due to its high computational efficiency and capability to handle large datasets effectively, providing robust performance with relatively low overfitting [68]. The preprocessing steps previously described were similarly utilized here, with the distinction that segregation energies are now incorporated as additional inputs, alongside the MC/MD states serving as the targets. The output layer is characterized by a logistic activation function, which predicts the occupancy state of an interface atom. The model's performance was assessed using the negative log-likelihood function as a performance metric, which aligns the predicted class probabilities with the true class distribution, thereby maximizing the likelihood of observing the correct class labels. XGBoost that was trained on the dataset utilized early stopping to mitigate overfitting. The model's performance was assessed using a confusion matrix and classification report on both training and test datasets, ensuring balance in target classes. Bayesian optimization [69] was applied to optimize parameters such as gamma (minimum loss reduction to split), maximum tree depth, learning rate, and minimum child weight (minimum sum of weights to split), followed by a grid search algorithm for fine-tuning the model parameters, as shown in Supplementary Note 4.

**Pipelines Tailored for Various Applications**

Fig. 3 displays four example configurations of the composable computational models. Figs. 3(a) and (b) depict pipelines previously developed for calculating site availabilities and DOSs in binary alloys [66] and for predicting site availabilities using ML techniques [44], respectively. For binary alloys, the A1 module represents a single-element base. An extension of these methodologies to CCAs is proposed in this work, which involves the use of multiple bases. As previously discussed, this necessitates defining reference states for each solute across different bases. Fig. 3(a) describes a pipeline based on direct simulations, employing an RSS simulation model to generate undecorated bases in their relaxed states. This determines the site availabilities and DOSs beyond the dilute limit. The pipeline in Fig. 3(b) introduces an ML approach, which is trained using site availabilities from atomistic simulations. The model then uses these predicted site availabilities to approximate segregation spectra and provide rapid and accurate predictions for site availabilities. To the best of the authors' knowledge, an ML pipeline for predicting DOSs in CCEs has not been previously attempted.

The pipeline shown in Fig. 3(c) addresses this key opportunity by applying the state matching procedure proposed in this study to determine MC/MD states that indicate segregation or negative segregation behavior. The binary nature of these states allows the problem to be framed as a classification task, suitable for ML techniques. This pipeline marks progress in predicting DOSs in CCEs without fully eliminating atomistic simulations for site availabilities. Advancing further, Fig. 3(d) illustrates a pipeline that seeks to obviate the need for atomistic simulations altogether. It employs ML to predict site availabilities, which are then combined with vectorized local atomic environments to predict DOSs in CCEs at any solute concentration. This pipeline indicates a substantial move towards an entirely ML-driven approach, enabling the exploration of segregation in CCEs without costly and time-consuming atomic scale simulations.

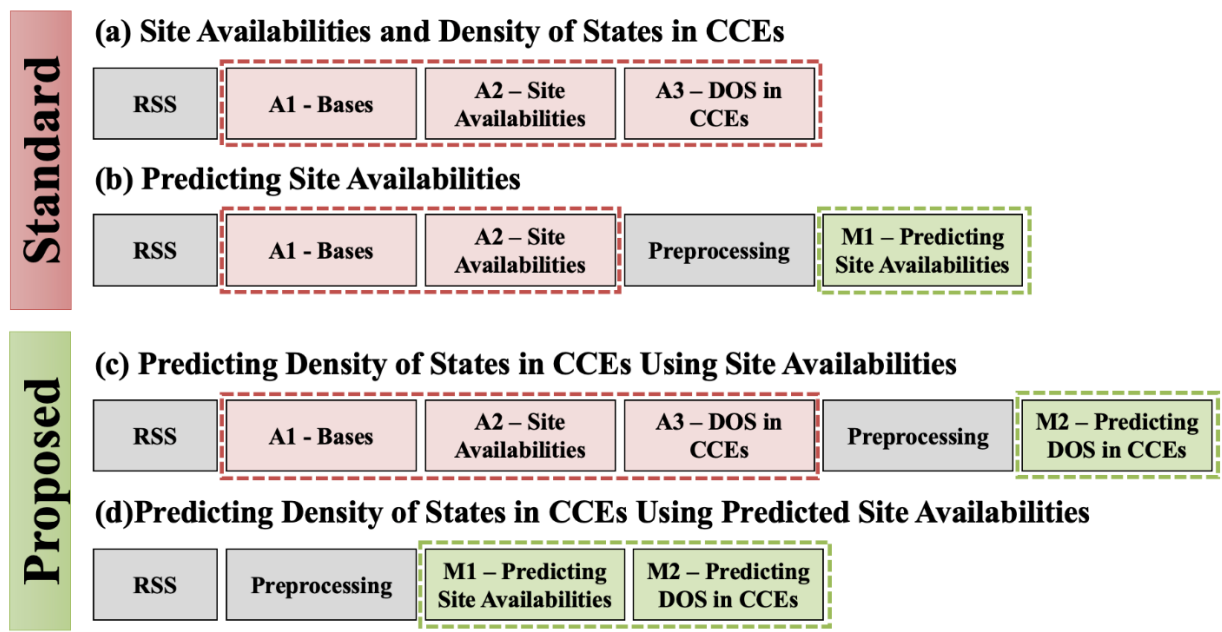

Fig. 3. Modeling pipelines tailored for different applications. Two pipelines previously developed for binary systems are designated as "Standard" yet are expanded for CCAs in this work: (a) The atomistic simulation pipeline for determining site availabilities at absolute zero and DOSs in chemically complex environments, and (b) the hybrid ML/atomistic simulation pipeline for predicting site availabilities. Proposed pipelines in this work (c) for predicting DOSs beyond the dilute limit, which utilizes a state matching procedure to determine instances of negative segregation, and (d) for predicting DOSs beyond the dilute limit from the predicted site availabilities, representing a significant step towards removing time-consuming atomistic simulations from the analysis. The dashed lines enclosing the modules represent the type of computational method used.

## 3. RESULTS AND DISCUSSIONS

**Site Availabilities at Absolute Zero**

In a multicomponent system, while some pair-interactions might seemingly average out, invoking a mean-field-like approximation, the actual methodology diverges by capturing individual A-B interactions, site-dependent behaviors, and the energy landscapes across different alloy compositions. The segregation spectra for each base-solute (A-B) type interaction (A as single element representing a binary interaction, and A as a multi-element alloy corresponding to ternary and quaternary interactions) in the NbMoTaW RCCA (obtained through the pipeline depicted in Fig. 3(a)) are shown in Fig. 4. These segregation energies will be referred to as "true" energies to signify that these were obtained from simulations, as opposed to the predicted energies obtained from ML models. A diverse array of segregation behaviors is demonstrated in Fig. 4, which highlights the necessity for an expanded approach in the predictive modeling of interfacial segregation within CCAs. Traditional models may be insufficient in capturing these complex behaviors since this requires identification of this behavior for the same solute in different bases. For example, the effect of CSRO is apparent in the case where Mo is the solute atom in Ta, $Nb_{0.5}Ta_{0.5}$, and $Nb_{0.33}Ta_{0.33}W_{0.34}$ bases, which are marked with red rectangles in Fig. 4. The binary case (Ta as base, Mo as solute) shows a distribution that consists of almost entirely anti-segregation sites, which could be originating from the strong cluster formation tendencies of Mo and Ta atoms structured as B2 clusters [5,70]. This tendency persists in Fig. 4(d) for the equiatomic MoTa base, where intermetallic formation in the bulk region is more pronounced than in the interface region. However, with the introduction of Nb to the system ($Nb_{0.5}Ta_{0.5}$ as base, Mo as solute), Mo segregation becomes more favorable. In the quaternary case ($Nb_{0.33}Ta_{0.33}W_{0.34}$ as base, Mo as solute), the inclusion of W atoms further pushes the distribution to left, resulting in a strong favorable segregation peak for Mo. For this RCCA, Ta-W interaction was identified to be attractive in the bulk and repulsive in the interface regions at room temperature [20], which aligns well with our observations here. In addition, Nb is more inclined to segregate to interfaces compared to W, which agrees with most available studies on the segregation behavior in NbMoTaW [4,5,70]. The broad spectra observed for ternaries and quaternaries also suggest that there are many more site-specific energy states that solute atoms can occupy, complicating the prediction of segregation behaviors. To translate these insights gained from the segregation energy spectra, the preprocessing module is utilized to prepare the data in a format amenable to ML model training.

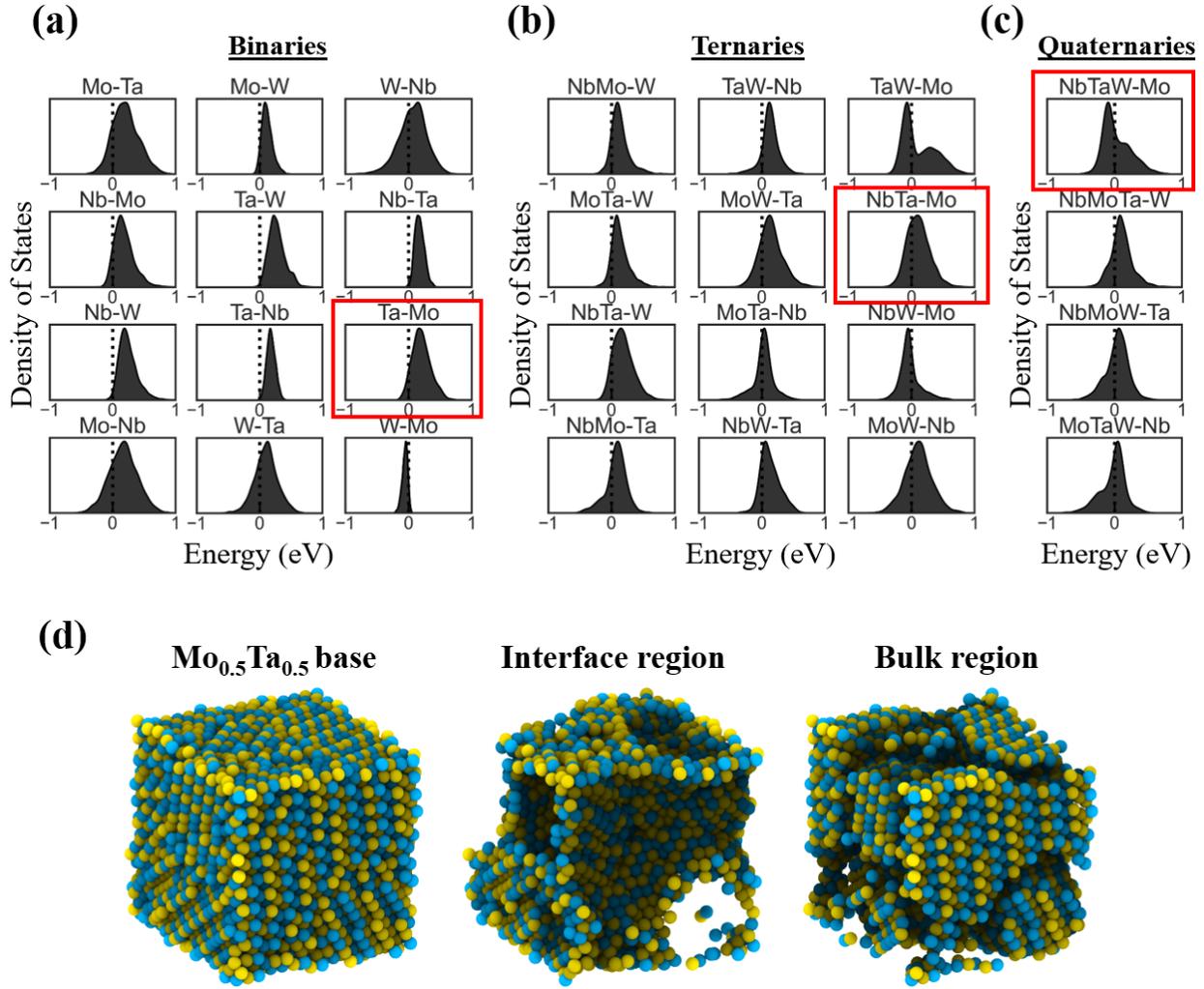

Fig. 4. Site availability data at absolute zero. Spectra of segregation energies corresponding to (a) binary (A as single element, e.g., Mo in Mo-Ta), (b) ternary (A comprising of two elements, e.g., $Nb_{0.5}Mo_{0.5}$ in NbMo-W), and (c) quaternary (A comprising of three elements, e.g., $Nb_{0.33}Ta_{0.33}W_{0.34}$ in NbTaW-Mo) interactions, and B (e.g., Ta in Mo-Ta, MoW-Ta, or NbMoW-Ta) is always a solute of a single element. The interface sites corresponding to negative and positive segregation energies are denoted as segregation and anti-segregation sites, respectively, demonstrating the variable segregation tendencies across different base configurations and solute combinations. The red rectangles guide the eye for the example system described in text. (d) The effect of Mo-Ta B2 clustering is observed to be more pronounced in the bulk regions than the interface regions for the MoTa base material.

The ANN in the M1 module can now be employed where the true segregation energy distributions are used as targets, in accordance with the pipeline shown in Fig. 3(b). The proficiency of the model is demonstrated in Fig. 5, which compares the predicted segregation energies with the true values for the validation dataset. The model, trained on the chemical interactions depicted in Fig. 4, has its predictions evaluated across different elemental interactions. When all interactions are considered (Fig. 5(a)), the model's predictions correspond closely to the true segregation energies, yielding a MSE of 0.01. Although MAE is not included in the training, it is observed as an additional metric that is physically easier to understand, with the model achieving a MAE of 0.07 eV. Given that the entire distribution of segregation energies spans ~1 eV, a MAE of 0.07 eV indicates a high level of precision in the model's predictions. This performance is also illustrated in Fig. 5(a) by observing that the error margin is narrower than the bin width of the histogram that represents the distribution of segregation energies. The model exhibits even better performance when it is limited to quaternary interactions (Fig. 5(b)), yielding an MSE of approximately $10^{-4}$ and achieving segregation energy predictions with a MAE of 0.04 eV. The MSE for this case is observed to be within the limits of error resolution maintained during training, indicating a highly precise predictive capability. The enhanced performance in the case of quaternary interactions could be attributed to the reduced complexity and more defined chemical interactions compared to more complex systems. Finally, Fig. 5(c) further displays the model's consistent performance across different specific solutes within the quaternary system, maintaining high accuracy throughout. The additional spread in the segregation energy predictions in Fig. 5(a) primarily originates from binary and ternary interactions, as detailed in Supplementary Note 5.

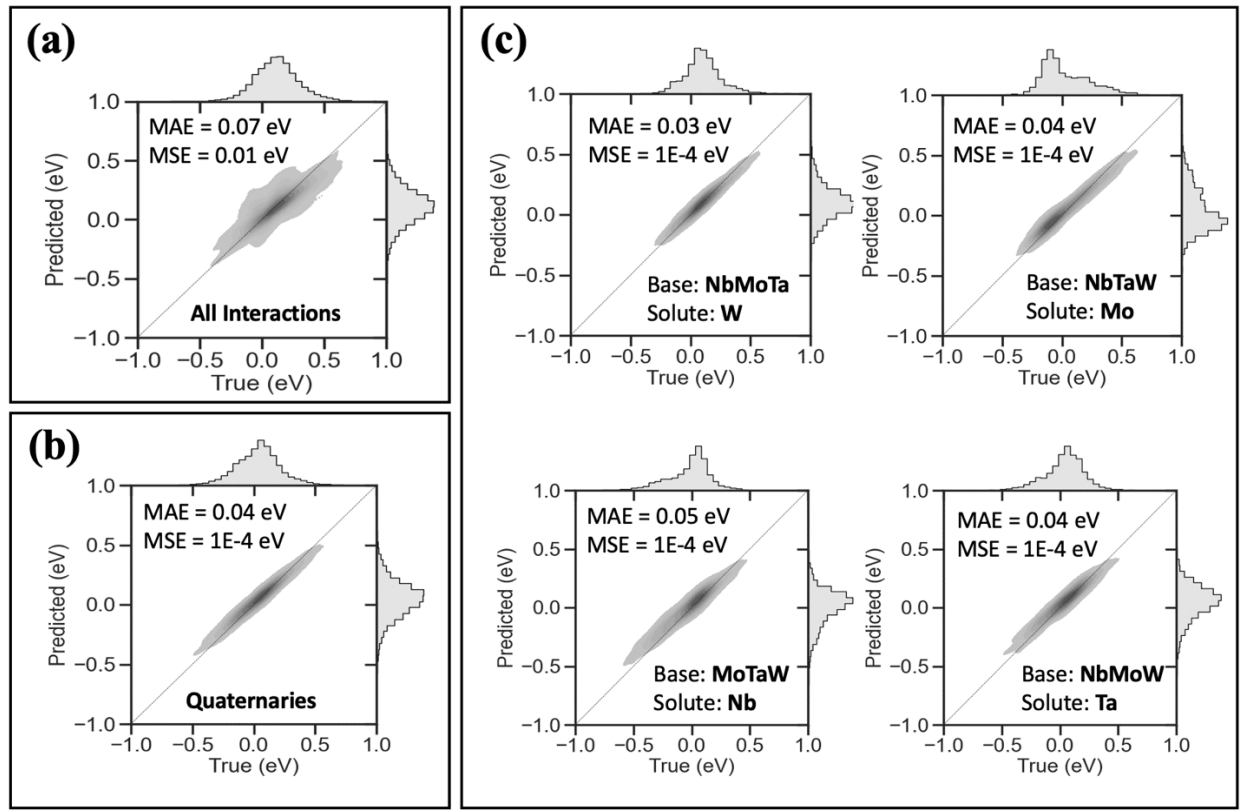

Fig. 5. Accuracy of neural network predictions for segregation energies. Comparison of the true segregation energies (calculated in module A2) with the predicted segregation energies for (a) all interactions, (b) only quaternary interactions, and (c) specific quaternary interactions with different solute atoms, confirming consistent high accuracy throughout. The mean absolute error (MAE) and mean squared error (MSE) values are shown on the upper left-hand side of each plot.

**Chemically Complex Environments**

With the inclusion of solute-solute interactions and thermal effects, the segregation landscape obtained in the previous section for zero temperature is no longer adequate for capturing the segregation behavior. To remedy this, the proposed pipeline in Fig. 3(c) is employed, which includes MC/MD simulations at 5 at.% solute-content increments, followed by the state matching procedure. Fig. 6(a) provides an example of the state matching process for a given base ($Nb_{0.5}Mo_{0.5}$), where base atoms are randomly replaced with Ta to achieve the desired composition of $Nb_{0.35}Mo_{0.35}Ta_{0.3}$. Following the MC/MD procedure, instances of negative segregation are revealed, where solute atoms in favorable interfacial sites are replaced by base atoms. In other words, not all energetically favorable sites are occupied due to the CCEs; some may instead favor anti-segregation sites, leaving these interface sites occupied by host atoms. One example of this behavior can be seen in Fig. 6(a), where the same interface site is marked with a red circle before and after equilibration. This site was occupied by a solute atom after equilibration, even though it had a positive segregation energy. This behavior becomes more common with increased solute concentration due to exhaustion of energetically favorable interface sites. The room temperature simulations mitigate temperature-induced structural transitions, such as interfacial disordering [20], although reaching this equilibrium for multielement systems is challenging due to their slow kinetics [3]. Fig. 6(b) shows the site occupancies for the same ternary system at room temperature, with solute concentrations adjusted in 10 at.% increments starting from 10 at.% concentration, and includes the corresponding site availability plot for reference. Fig. 6(c) shows the solute concentrations adjusted in 10 at.% increments starting from 5 at.% concentration in three separate plots for more discernible visualization of each distribution's characteristics. Plots corresponding to additional binary, ternary, and quaternary interactions can be found in Supplementary Note 6.

Consistent with prior literature, binary systems exhibit a distinct pattern of solute behavior. Initially, the distribution is narrow and focused towards the left, indicating the occupancy of energetically favorable (negative energy) segregation sites [22]. However, with increasing solute concentration, this distribution broadens and shifts to include some higher-energy (positive energy, anti-segregation) sites [66]. In assessing the MC/MD model's ability to predict segregation behaviors across varying solute concentrations, a key observation is made in the behavior of the DOS distributions. For ternary and quaternary systems, as the solute concentration increases, the DOS distribution retains its broad character, contrary to the expectation of a narrowing trend [66]. This broad, consistent shape persists even as the solute content rises. This difference in behavior highlights a complex segregation dynamic in CCEs as compared to simpler binary systems. For example, in the ternary NbMo-Ta system illustrated in Fig. 6(a) there is a broad energy distribution from the outset. This suggests that the presence of Nb, which is the strongest segregating species in any system containing it, restricts Ta's ability to occupy energetically favorable sites, thus maintaining a wide range of energy states even at low solute concentrations.

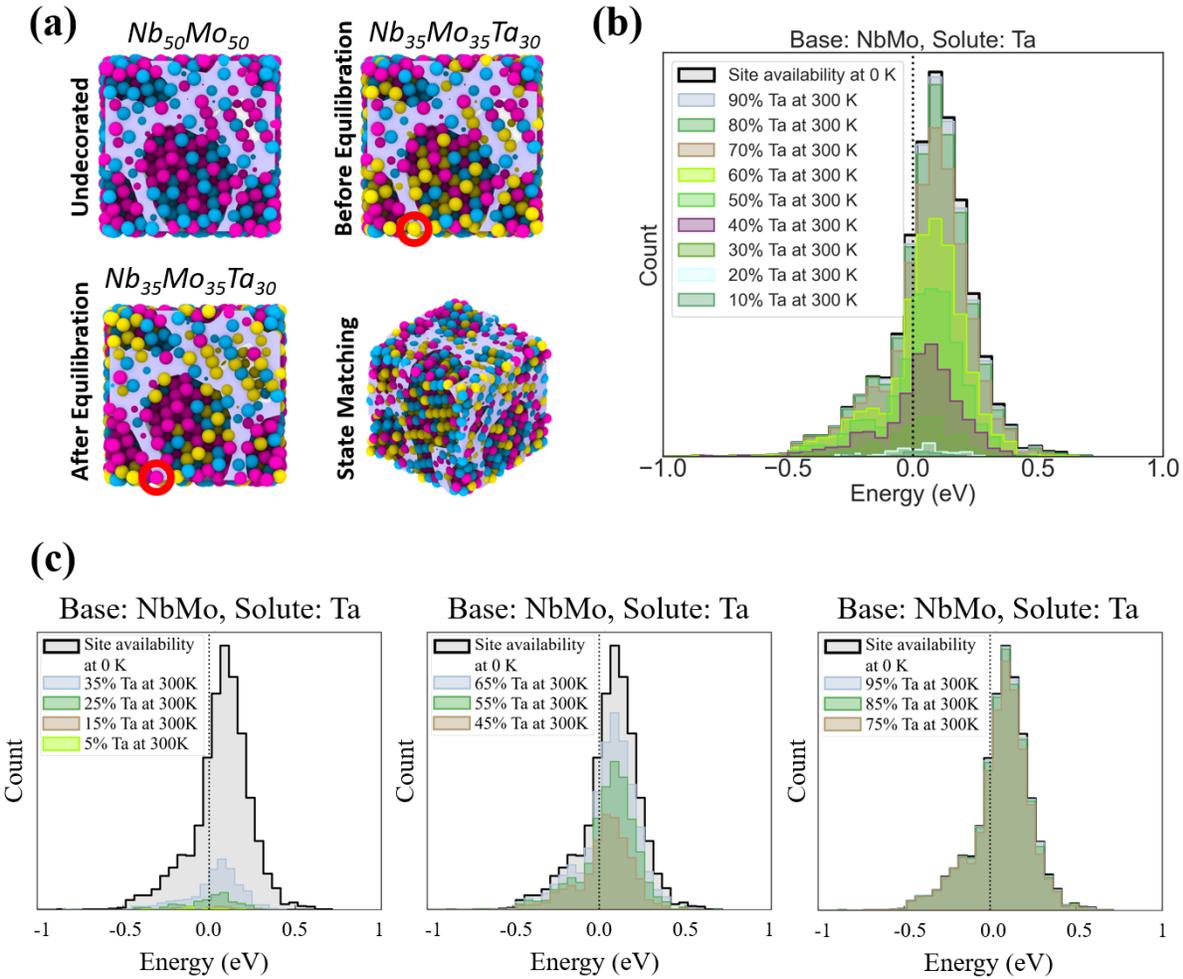

**Fig. 6. Visualization of solute distribution and segregation in chemically complex environments. (a) Illustration for a sample undecorated NbMo base demonstrating the state matching process after Ta substitution and**

**subsequent equilibration via MC/MD simulations, revealing solute atom distribution pre- and post-state matching. Gray regions correspond to the interfaces, while Nb, Mo, and Ta atoms are represented as pink, blue, and yellow atoms, respectively. The DOS at room temperature for a NbMo base with Ta as the solute, showing site occupancies at 10 at.% solute concentration increments starting from (b) 10 at.%, and (c) 5 at.%, and the tendency for negative segregation. The distributions in (c) are shown as three separate plots to mitigate visual overlap and improve interpretability.**

To better understand the effect of increased solute concentration on the predictive accuracy of the model, separate models were trained for different chemical interactions and varying solute concentrations. Fig. 7(a) illustrates the variation in test dataset accuracies, indicative of the model's efficacy in predicting MC/MD states across varying solute concentrations. Test accuracies are presented for models trained exclusively on binary, ternary, quaternary, and all interactions. Each concentration in Fig. 7(a) encompasses all lower concentrations within the training set; hence, a model trained for a 20 at.% solute composition is also trained on datasets representing 5 at.%, 10 at.%, and 15 at.% solute compositions. While the focus of the work is on CCEs, it is notable that simpler segregation behavior such as those of binary alloys can also come from this model. Notably, the model trained solely on binary interactions shows exceptional performance at lower solute concentrations, achieving an accuracy of approximately 96% in the test dataset.

The model trained on all interactions, which can predict the DOSs in CCEs, demonstrates a robust accuracy rate, remaining around 90% for the test dataset for solute concentrations up to the equiatomic concentration. This indicates that the DOSs of any ternary or higher CCA with solute content ranging from zero to near equiatomic concentrations can be predicted with a high level of accuracy at room temperature. For the binary system (Fig. 7(a)), such configurations do not have physical meaning and the graph terminates at 50 at.% due to the base-solute interchangeability at higher concentrations. This same symmetry doesn't exist for ternaries and quaternaries, as the multitude of A-B pair combinations blurs the distinction between base and solute beyond equiatomic levels, resulting in an equivalent elemental combination but with a non-equiatomic base. Beyond this, the uneven distribution of solute atoms can cause significant lattice distortions due to size mismatches and differing electrochemical properties [71–74]. Such distortions may lead to the emergence of new phases that were not accounted for in the initial model formulations [4,75,76], reducing the models' predictive capability. Moreover, the emergence of structural defects, such as vacancies and interstitials, can further complicate the material's behavior by acting as scattering centers and altering its intrinsic properties [77]. These combined factors result in deviations from the predictions of our model.

To provide deeper insight into the prediction dynamics, a confusion matrix for the test dataset is constructed for the all-interaction model up to the equiatomic concentration, as shown in Fig. 7(b). This matrix details the outcomes for the binary classification of MC/MD states, revealing a substantial distinction in the model's ability to accurately identify negative segregation versus segregation states. The matrix shows that 64,230 out of 66,088 instances are correctly predicted as negative segregation (state '0'), corresponding to a ~97% accuracy. In contrast, 4,637 instances are accurately identified as segregation (state '1') from a total of 10,248, yielding ~45% accuracy. The high true negative rate indicates that the model is highly effective at correctly identifying instances where negative segregation occurs, suggesting that the model is well-tuned to recognize stable configurations where solutes remain distributed within the bulk matrix. However, the true positive rate is markedly lower. True positives in this context refer to the model's ability to accurately predict when a solute atom will segregate to an interface site. The relatively low true positive rate indicates that the model is less adept at predicting when solute atoms preferentially occupy interface sites, which could lead to underestimating the alloy's propensity for segregation-related phenomena such as embrittlement or enhanced chemical reactivity at the interfaces.

An important factor in this context is the impact of MC/MD parameters and sampling on the identification of negative segregation and segregation states, which could influence atom tracking due to thermal and temporal fluctuations, potentially affecting the identification of grain structure and subsequent segregation states. Such influence is complicated by the energetic near-degeneracy of many chemical environments in CCAs [71]. As an example, Mo interfacial atomic fluctuations in a NbTaW base with Mo solute (30%) were measured at 1.76% over the last 200 timesteps of a 1000 timestep MC/MD run. This fluctuation range highlights the susceptibility to variations. Additionally, the adaptive common neighbor analysis method's vulnerability to thermal fluctuations needs to be noted, which could result in under- or over-identification of segregation states. Given the energy-driven nature of the simulations, it is noted that final configurations should stabilize similarly; however, extending the simulations or optimizing the number of configurations sampled in MC/MD are proposed strategies to address these potential discrepancies.

The imbalance in the dataset, with a much higher count of true negatives compared to true positives, indicates that negative segregation is more prevalent than segregation in the test dataset. This skew could be attributed to solute-solute interactions [66] as well as to the high-solubility limits of CCAs [2]. Addressing the imbalance in the dataset by incorporating more instances of segregation, which involves changing the bulk-to-interface ratio, or changing the model size, and refining the features used to represent the atomic environments to mitigate its conservative bias are potential strategies for future improvement of ML models of the type introduced here. The false negative rate of 7.35% implies that the model occasionally misses instances of segregation, meaning that the model might fail to capture some interactions leading to segregation at the interface. On the other hand, the false positive rate of 2.43% suggests that the model sometimes incorrectly predicts segregation where it does not occur, which may reflect an inherent bias in the model due to the imbalanced nature of segregation compared to negative segregation in this CCA. Overall, the higher rate of false negatives compared to false positives suggests that the model is more conservative in predicting segregation. This tendency is exacerbated with increased solute concentration, indicating that predicting the segregation behavior gets more challenging as the local atomic environments become more complex due to anti-segregation sites starting to fill up. Nevertheless, these results highlight the model's strong predictive power in identifying and categorizing interfacial segregation behavior in chemically complex environments.

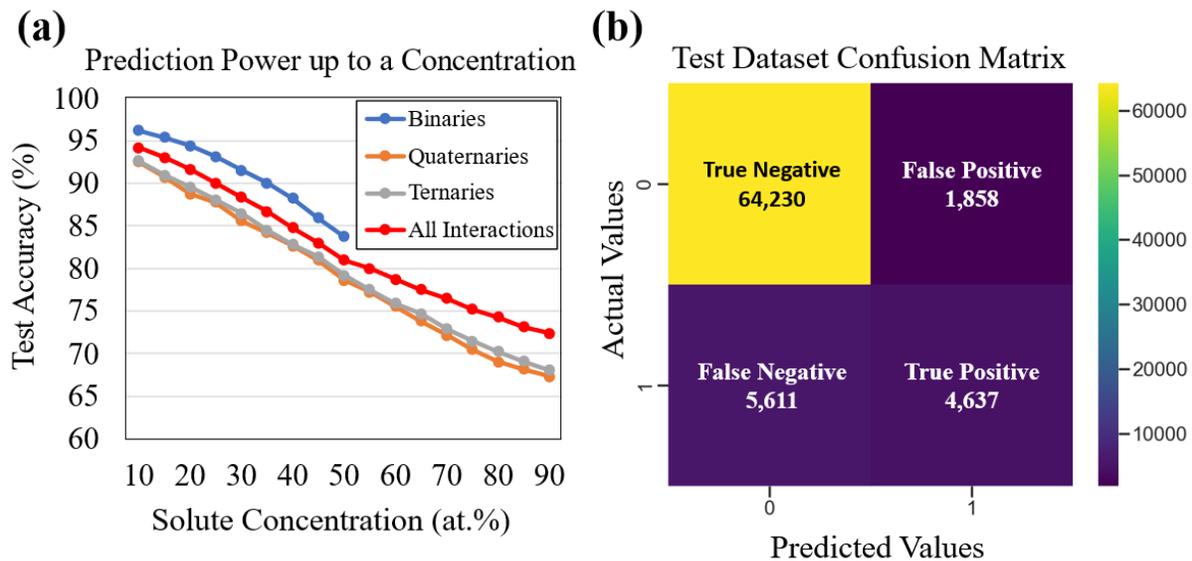

**Fig. 7. Classification model performance and prediction accuracy for varied solute concentration. (a) The test accuracy variation with increasing solute concentrations for binary, ternary, quaternary, and all interaction models. The model retains high accuracy up to the equiatomic concentration. (b) The confusion matrix for the all-interaction model, emphasizing the model's efficacy in predicting negative segregation and segregation states at interface sites.**

**Predicting Negative Segregation Behavior**

The computational overhead of atomistic simulations performed for obtaining the negative segregation behavior in CCEs is significant, even for this relatively small model. For example, the atomistic simulations in modules A1-A3 require around 72 days of CPU time using 64 cores. In contrast, using 12 cores, M1 predictions only take about 2 minutes while M2 predictions take less than a second. Although the number of cores does not scale linearly with simulation time, this roughly means that simulating a new composition using the prediction pipeline shown in Fig. 2(d) instead of atomistic simulations will take minutes instead of months in CPU time.

The approach described here employs the pre-trained M1 model to make predictions of site availabilities, which were shown to be highly accurate, and passes these predictions to the M2 model, along with the vectorized local atomic environments. Therefore, once M1 and M2 models are trained for a material system, this pipeline can provide very fast predictions of the DOSs in CCEs. Such an approach is not only orders of magnitude faster than conventional methods but also maintains high accuracy. As depicted in the confusion matrix shown in Fig. 8(a), the model, which utilizes predicted site availabilities, exhibits an 87% accuracy for the test dataset for solute concentrations up to the equiatomic concentration. This represents only a 3% decrease in accuracy compared to running more computationally costly atomistic simulations. Moreover, the confusion matrix suggests that while the inherent bias towards negative segregation states is preserved in this approach, the relative importance of the false negative rates is amplified.

Additionally, the interpretable nature of the XGBoost algorithm, unlike ANN methods, facilitates deeper insights [68]. This is exemplified by the ability to generate estimates of feature importance plots directly from the model, as shown in Fig. 8(b). These plots provide valuable assessments of the importance of local atomic environments relative to thermodynamic considerations in determining segregation behavior in CCEs. Fig. 8(b) shows the F-score, which is a statistical measure combining both precision and recall evaluating the accuracy of a model's predictions, of individual features used in the training process. According to this plot, segregation energies are the most important feature, indicating that they have the greatest influence on the model's ability to predict segregation behavior accurately in the given dataset. The other features include dimensionality-reduced SOAP vector representations. These vectors represent local atomic environments using a mathematical framework based on spherical harmonics, which, while effective for machine learning applications, do not provide a direct physical mapping of the atomic arrangements [63].

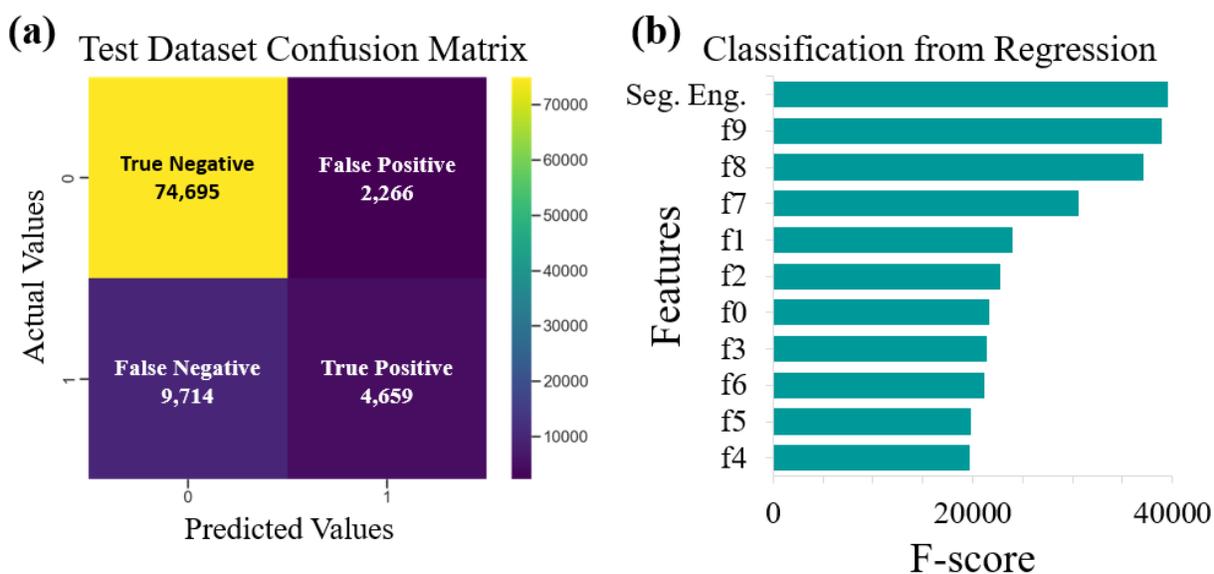

Fig. 8. Performance metrics and feature importance for segregation prediction. (a) The confusion matrix for a model using predicted site availabilities, with an 87% accuracy rate for solute concentrations up to the equiatomic concentration. The matrix highlights the model's inherent bias towards negative segregation. (b) The F-scores for features used in XGBoost training, emphasizing segregation energies as the most influential feature for accurate predictions in CCAs.

**Summary and Conclusions**

The ML framework described here, aimed at predicting interfacial segregation behavior in chemically complex environments, resonates with emerging trends in materials science, where accurate predictions often hinge on the careful selection and understanding of features. For instance, selecting physically meaningful features is especially important when there is limited data availability [37]. Similarly, the advancements in CCA research demonstrate the impact of comprehensive feature engineering in enhancing prediction accuracy of phases with small datasets [35]. Finally, the use of ANNs in the study of grain boundary property diagrams illustrates the potential of ML to simplify and effectively analyze complex, multidimensional problems, similar to the challenges in modeling CCEs [33]. These studies collectively reinforce the approach taken in our study, where feature importance not only guides the prediction accuracy but also deepens our understanding of the interfacial segregation phenomena in CCEs. Moreover, the scope of this pipeline extends beyond its current application, as it can be adapted for a variety of material systems and modified to account for segregation behavior across different temperature ranges.

In summary, the robust and adaptable tool developed in this study offers a means for researchers to conduct high-throughput investigations into co-segregation behavior in CCEs, eliminating the need for expensive atomistic simulations. It thus serves as a valuable asset in elucidating the complex chemical ordering and clustering behavior that is characteristic of CCAs.

## 4. DATA AVAILABILITY

All data generated or analyzed during this study are included in this published article and its supplementary information files.

## 5. CODE AVAILABILITY

The underlying code for this study will be made available to researchers on reasonable request from the corresponding author.

## 6. ACKNOWLEDGEMENTS

This research was primarily supported by the National Science Foundation Materials Research Science and Engineering Center program through the UC Irvine Center for Complex and Active Materials (DMR-2011967). In the preparation of this article, a large language model (ChatGPT September 25, 2023) was employed for tasks such as proofreading, code optimization, and code commentary through careful prompt engineering. While the tool aided in these areas, it was not involved in the conceptualization of the work. All outputs from ChatGPT were rigorously examined for factual accuracy.

## 7. AUTHOR CONTRIBUTIONS

**DA**: Formal Analysis; Atomistic Simulations; Software Development; Methodology; Investigation; Writing - original draft; Writing - review & editing. **JL**: Conceptualization; Writing - review & editing. **PC**: Conceptualization; Writing - review & editing. **TJR**: Conceptualization; Supervision; Funding acquisition; Project Administration; Writing - review & editing.

## 8. CONFLICT OF INTEREST

All authors declare no financial or non-financial competing interests.

# 9. REFERENCES

[1] M.A. Tunes, V.M. Vishnyakov, Materials & Design 170 (2019) 107692.
[2] O.N. Senkov, G.B. Wilks, D.B. Miracle, C.P. Chuang, P.K. Liaw, Intermetallics 18 (2010) 1758–1765.
[3] P. Cao, Acc. Mater. Res. 2 (2021) 71–74.
[4] T. Kostiuchenko, F. Körmann, J. Neugebauer, A. Shapeev, Npj Comput Mater 5 (2019) 55.
[5] J. Byggmästar, K. Nordlund, F. Djurabekova, Phys. Rev. B 104 (2021) 104101.
[6] R. Kozak, A. Sologubenko, W. Steurer, Zeitschrift Für Kristallographie - Crystalline Materials 230 (2015) 55–68.
[7] O.N. Senkov, G.B. Wilks, J.M. Scott, D.B. Miracle, Intermetallics 19 (2011) 698–706.
[8] D. Chatain, P. Wynblatt, Computational Materials Science 187 (2021) 110101.
[9] Z.D. Han, H.W. Luan, X. Liu, N. Chen, X.Y. Li, Y. Shao, K.F. Yao, Materials Science and Engineering: A 712 (2018) 380–385.
[10] X.B. Feng, J.Y. Zhang, Y.Q. Wang, Z.Q. Hou, K. Wu, G. Liu, J. Sun, International Journal of Plasticity 95 (2017) 264–277.
[11] D. Farkas, J Mater Sci 55 (2020) 9173–9183.
[12] S. Chen, Z.H. Aitken, S. Pattamatta, Z. Wu, Z.G. Yu, D.J. Srolovitz, P.K. Liaw, Y.-W. Zhang, Nat Commun 12 (2021) 4953.
[13] Q.F. He, P.H. Tang, H.A. Chen, S. Lan, J.G. Wang, J.H. Luan, M. Du, Y. Liu, C.T. Liu, C.W. Pao, Y. Yang, Acta Materialia 216 (2021) 117140.
[14] P. Garg, Z. Pan, V. Turlo, T.J. Rupert, Acta Materialia 218 (2021) 117213.
[15] D. Aksoy, R. Dingreville, D.E. Spearot, Acta Materialia 205 (2021).
[16] D. Aksoy, R. Dingreville, D.E. Spearot, Modelling and Simulation in Materials Science and Engineering 27 (2019).
[17] D. Scheiber, L. Romaner, Acta Materialia 221 (2021) 117393.
[18] R. Dingreville, D. Aksoy, D.E. Spearot, Scientific Reports 7 (2017).
[19] M.J. McCarthy, H. Zheng, D. Apelian, W.J. Bowman, H. Hahn, J. Luo, S.P. Ong, X. Pan, T.J. Rupert, Phys. Rev. Materials 5 (2021) 113601.
[20] D. Aksoy, M.J. McCarthy, I. Geiger, D. Apelian, H. Hahn, E.J. Lavernia, J. Luo, H. Xin, T.J. Rupert, Journal of Applied Physics 132 (2022).
[21] F. Körmann, A.V. Ruban, M.H.F. Sluiter, Materials Research Letters 5 (2017) 35–40.
[22] M. Wagih, C.A. Schuh, Acta Materialia 181 (2019) 228–237.
[23] S.Y. Lee, S. Byeon, H.S. Kim, H. Jin, S. Lee, Materials & Design 197 (2021) 109260.
[24] G. Deffrennes, K. Terayama, T. Abe, R. Tamura, Materials & Design 215 (2022) 110497.
[25] Y.V. Krishna, U.K. Jaiswal, R.M. R, Scripta Materialia 197 (2021) 113804.
[26] W. Huang, P. Martin, H.L. Zhuang, Acta Materialia 169 (2019) 225–236.
[27] K. Lee, M.V. Ayyasamy, P. Delsa, T.Q. Hartnett, P.V. Balachandran, Npj Comput Mater 8 (2022) 25.
[28] Q. Han, Z. Lu, S. Zhao, Y. Su, H. Cui, Computational Materials Science 215 (2022) 111774.
[29] Y. Zhang, C. Wen, C. Wang, S. Antonov, D. Xue, Y. Bai, Y. Su, Acta Materialia 185 (2020) 528–539.
[30] M. Dai, M.F. Demirel, Y. Liang, J.-M. Hu, Npj Comput Mater 7 (2021) 103.
[31] S. Hou, M. Sun, M. Bai, D. Lin, Y. Li, W. Liu, Acta Materialia 228 (2022) 117742.

[32] X. Hu, J. Wang, Y. Wang, J. Li, Z. Wang, Y. Dang, Y. Gu, Computational Materials Science 155 (2018) 331–339.
[33] C. Hu, Y. Zuo, C. Chen, S. Ping Ong, J. Luo, Materials Today 38 (2020) 49–57.
[34] C. Chen, S.P. Ong, Npj Comput Mater 7 (2021) 173.
[35] D. Dai, T. Xu, X. Wei, G. Ding, Y. Xu, J. Zhang, H. Zhang, Computational Materials Science 175 (2020) 109618.
[36] M. Guziewski, D. Montes De Oca Zapiain, R. Dingreville, S.P. Coleman, ACS Appl. Mater. Interfaces 13 (2021) 3311–3324.
[37] P.-P. De Breuck, G. Hautier, G.-M. Rignanese, Npj Comput Mater 7 (2021) 83.
[38] Z. Pei, J. Yin, J.A. Hawk, D.E. Alman, M.C. Gao, Npj Comput Mater 6 (2020) 50.
[39] Z. Zhou, Y. Zhou, Q. He, Z. Ding, F. Li, Y. Yang, Npj Comput Mater 5 (2019) 128.
[40] S. Yang, N. Zhou, H. Zheng, S.P. Ong, J. Luo, Phys. Rev. Lett. 120 (2018) 085702.
[41] T. Frolov, W. Setyawan, R.J. Kurtz, J. Marian, A.R. Oganov, R.E. Rudd, Q. Zhu, Nanoscale 10 (2018) 8253–8268.
[42] J. Han, V. Vitek, D.J. Srolovitz, Acta Materialia 104 (2016) 259–273.
[43] W.F. Reinhart, Computational Materials Science 196 (2021) 110511.
[44] L. Huber, R. Hadian, B. Grabowski, J. Neugebauer, Npj Comput Mater 4 (2018) 64.
[45] M. Wagih, P.M. Larsen, C.A. Schuh, Nat Commun 11 (2020) 6376.
[46] W. Dai, H. Wang, Q. Guan, D. Li, Y. Peng, C.N. Tomé, Acta Materialia 214 (2021) 117006.
[47] X. Liu, J. Zhang, Z. Pei, Progress in Materials Science 131 (2023) 101018.
[48] W. Ye, H. Zheng, C. Chen, S.P. Ong, Scripta Materialia 218 (2022) 114803.
[49] D. McLean, A. Maradudin, Physics Today 11 (1958) 35–36.
[50] C.L. White, W.A. Coghlan, Metall Trans A 8 (1977) 1403–1412.
[51] P. Hirel, Computer Physics Communications 197 (2015) 212–219.
[52] N. Tuchinda, C.A. Schuh, Acta Materialia 226 (2022) 117614.
[53] A.P. Thompson, H.M. Aktulga, R. Berger, D.S. Bolintineanu, W.M. Brown, P.S. Crozier, P.J. In 'T Veld, A. Kohlmeyer, S.G. Moore, T.D. Nguyen, R. Shan, M.J. Stevens, J. Tranchida, C. Trott, S.J. Plimpton, Computer Physics Communications 271 (2022) 108171.
[54] A. Stukowski, Modelling and Simulation in Materials Science and Engineering 18 (2010).
[55] T.A. Caswell, M. Droettboom, A. Lee, E.S. de Andrade, T. Hoffmann, J. Hunter, J. Klymak, E. Firing, D. Stansby, N. Varoquaux, J.H. Nielsen, B. Root, R. May, P. Elson, J.K. Seppänen, D. Dale, J.-J. Lee, D. McDougall, A. Straw, P. Hobson, hannah, C. Gohlke, A.F. Vincent, T.S. Yu, E. Ma, S. Silvester, C. Moad, N. Kniazev, E. Ernest, P. Ivanov, (2021).
[56] M.L. Waskom, Journal of Open Source Software 6 (2021) 3021.
[57] S. Yin, Y. Zuo, A. Abu-Odeh, H. Zheng, X.-G. Li, J. Ding, S.P. Ong, M. Asta, R.O. Ritchie, Nat Commun 12 (2021) 4873.
[58] Y. Zuo, C. Chen, X. Li, Z. Deng, Y. Chen, J. Behler, G. Csányi, A.V. Shapeev, A.P. Thompson, M.A. Wood, S.P. Ong, J. Phys. Chem. A 124 (2020) 731–745.
[59] S. Yin, J. Ding, M. Asta, R.O. Ritchie, Npj Comput Mater 6 (2020) 1–11.
[60] A.V. Shapeev, Multiscale Model. Simul. 14 (2016) 1153–1173.
[61] J. Behler, The Journal of Chemical Physics 134 (2011) 074106.
[62] J. Behler, M. Parrinello, Phys. Rev. Lett. 98 (2007) 146401.
[63] C.W. Rosenbrock, E.R. Homer, G. Csányi, G.L.W. Hart, Npj Comput Mater 3 (2017) 29.
[64] G. Csanyi, S. Winfield, J. Kermode, M.C. Payne, A. Comisso, A. De Vita, N. Bernstein, Newsletter of the Computational Physics Group (2007) 1–24.
[65] J.R. Kermode, J. Phys.: Condens. Matter 32 (2020) 305901.

[66]     M. Wagih, C.A. Schuh, Acta Materialia 217 (2021) 117177.
[67]     A. Botchkarev, IJIKM 14 (2019) 045–076.
[68]     T. Chen, C. Guestrin, in: Proceedings of the 22nd ACM SIGKDD International Conference on Knowledge Discovery and Data Mining, 2016, pp. 785–794.
[69]     J. Snoek, H. Larochelle, R.P. Adams, (2012).
[70]     X. Liu, J. Zhang, J. Yin, S. Bi, M. Eisenbach, Y. Wang, Computational Materials Science 187 (2021) 110135.
[71]     L.R. Owen, N.G. Jones, J. Mater. Res. 33 (2018) 2954–2969.
[72]     D.B. Miracle, O.N. Senkov, Acta Materialia 122 (2017) 448–511.
[73]     Z. Wang, W. Qiu, Y. Yang, C.T. Liu, Intermetallics 64 (2015) 63–69.
[74]     E. Blokker, X. Sun, J. Poater, J.M. Van Der Schuur, T.A. Hamlin, F.M. Bickelhaupt, Chemistry A European J 27 (2021) 15616–15622.
[75]     N. Zhou, T. Hu, J. Huang, J. Luo, Scripta Materialia 124 (2016) 160–163.
[76]     F. Körmann, M. Sluiter, Entropy 18 (2016) 403.
[77]     D. Aksoy, P. Cao, J.R. Trelewicz, J.P. Wharry, T.J. Rupert, JOM (2024).